\documentclass[aps,amsfonts,amssymb, amsmath, prl, ]{revtex4}

\newcommand{\RR}{{\mathbb R}}
\newcommand{\ra}{\rightarrow}
\newcommand{\eps}{\epsilon}

\newcommand{\dist}{{\rm dist}}
\newcommand{\supp}{{\rm supp}}

\begin{document}

\title{Absence of energy currents in an equilibrium state and chiral anomalies}

\author{Anton Kapustin}
\email{kapustin@theory.caltech.edu}
\author{Lev Spodyneiko}
\email{lionspo@caltech.edu}
\affiliation{California Institute of Technology, Pasadena, CA 91125, United States}

\begin{abstract}
A long time ago F. Bloch showed that in a system of interacting non-relativistic particles the net particle-number current must vanish in any equilibrium state. Bloch's argument does not generalize easily to the energy current. We devise an alternative argument which proves the vanishing of the net energy currents in equilibrium states of lattice systems as well as  systems of non-relativistic particles with finite-range potential interactions. We discuss some applications of these results. In particular, we show that neither a 1d lattice system nor a 1d system of non-relativistic particles with finite-range potential interactions can flow to a Conformal Field Theory with unequal left-moving and right-moving central charges.
\end{abstract}

\maketitle

\paragraph{Introduction}
An old argument of F. Bloch explained in detail by D. Bohm \cite{Bohm}  shows that in an equilibrium state of a quasi-1d system of non-relativistic particles the net particle number current is zero. By an equilibrium  state we mean either a ground state or a Gibbs state. A quasi-1d system is a system which is infinitely-extended in only one direction, but can have an arbitrary number of finite directions. Recently H. Watanabe extended Bloch's argument to lattice systems \cite{Watanabe}.  This result appears very general and likely to apply to currents of other conserved quantities. For example, a non-vanishing energy current in an equilibrium state would conflict with  Fourier's law. However, the Bloch-Bohm argument does not immediately apply to the energy current, since it relies in an essential way on the quantization of the particle number which does not have an analog in the case of energy.

There are also examples of systems where equilibrium currents do not vanish. In a 1+1d Conformal Field Theory with unequal central charges $c_L$ and $c_R$ for left-movers and right-movers the energy current at temperature $T$ is nonzero end equal to \cite{CFT1,CFT2}
\begin{equation}\label{chiralc}
\langle j_E\rangle=\frac{\pi T^2}{12}\left(c_R-c_L\right).
\end{equation}
In a 1+1d CFT with a $U(1)$ current algebra at levels $k_L$ and $k_R$  the net $U(1)$ current at a chemical potential $\mu$ and arbitrary temperature is \cite{muCFT}
\begin{equation}\label{chiralk}
\langle j_Q\rangle=\frac{\mu}{2\pi} (k_R-k_L).
\end{equation}
This raises the question about the precise conditions under which equilibrium-state currents vanish. Note that in both examples symmetry anomalies (i.e. obstructions to gauging a symmetry) are present: $k_R-k_L$ measures the anomaly of the $U(1)$ symmetry, while $c_R-c_L$ measures the anomaly of the diffeomorphism symmetry. In the case of the $U(1)$ symmetry, this implies that a lattice system with an on-site $U(1)$ symmetry or a system of non-relativistic particles cannot flow to a CFT with a  non-zero $k_R-k_L$. Indeed, if a system can be consistently coupled to a $U(1)$ gauge field, the same must hold for its long-wavelength limit, ruling out a  low-energy theory with a nontrivial $U(1)$ anomaly. This argument does not work for energy currents since most microscopic Hamiltonians in condensed matter physics  cannot be coupled to a gravity in any natural way. To linear order, a natural coupling to gravity requires a conserved symmetric stress-energy tensor. In relativistic field theory, such a tensor is present because the system is invariant under continuous  translations in space and time and Lorenz transformations, but most microscopic models in condensed matter physics are invariant only under continuous time translations and discrete spatial translations. Nevertheless, it is widely believed that a system of particles with short-range interactions or a lattice system with short-range interactions  cannot flow to a CFT with a non-zero $c_R-c_L$. 

In this letter, we prove the absence of equilibrium-state energy  currents for quasi-1d lattice systems with  finite-range interactions as well as for systems of non-relativistic particles with finite-range potential interactions. We make only very modest assumptions, which roughly amount to the absence of phase transitions in quasi-1d systems at positive temperatures.\footnote{Many-body localization transitions are not accompanied by divergent  susceptibilities and are not regarded as phase transitions for our purposes.} Our arguments apply equally to bosons and fermions. %An outline of the argument for lattice systems is as follows. Let $x$ be the coordinate in the infinite direction. Consider an infinitesimal deformation of the Hamiltonian which vanishes for $x>0$. The effect of such a deformation on the expectation value of any local observable supported at $x=R$, where $R$ is large and positive, should be small away from phase transitions, since the linear response kernels decay rapidly. But in a stationary state the current expectation value is independent of position. By taking $R$ to infinity, we conclude that the expectation value of the current is unchanged everywhere, including $x<0$. Now we can decompose an arbitrary deformation of the Hamiltonian into one supported at $x< \delta$ and one supported at $x>-\delta$. The above argument plus the  linearity of response then imply that the current expectation value is unchanged under arbitrary deformations of the Hamiltonian. Thus we can compute the expectation value of the current by deforming the Hamiltonian to a simpler one. For example, we can deform the Hamiltonian to zero, or equivalently increase the temperature to infinity. For $T=\infty$ it is easy to see that the energy current expectation value vanishes, hence the result follows. 
An immediate corollary is that lattice systems or systems of particles with  finite-range interactions cannot flow to a 1+1d CFT with a non-zero $c_R-c_L$. We also give an alternative derivation of the vanishing of the $U(1)$ current in certain continuous systems and explain how 1+1d chiral CFTs perturbed by a chemical potential manage to evade this conclusion. 

A. K. is grateful to H. Watanabe for a discussion of Bloch's theorem.
This research was supported in part by the U.S.\ Department of Energy, Office of Science, Office of High Energy Physics, under Award Number DE-SC0011632. A.\ K.\ was also supported by the Simons Investigator Award. 

\paragraph{Lattice systems}

%A lattice system in $d$ spatial dimensions has a Hilbert space $V=\otimes_{p\in\Lambda} V_p$, where $\Lambda$ (``the lattice'') is a discrete subset of $\RR^d$, and $V_p$ is finite-dimensional. An observable is  localized at a point $p\in\Lambda$ if it has the form $A\otimes 1_{\Lambda\backslash p}$ for some $A: V_p\ra V_p$. An observable is localized at a subset $Q\subset\Lambda$ if it commutes with all observables localized at any  $p\in \Lambda\backslash Q$. For a compact $Q$, this implies that an observable localized at $Q$ is a sum of products of observables localized at all $p\in Q$. An observable localized in a compact set $Q$ will be called a local observable with support $Q$. 

The Hamiltonian of a lattice system has the form 
\begin{equation}
H=\sum_{p\in\Lambda} H_p,
\end{equation}
where $\Lambda\subset\RR^n$ is the lattice.
We assume that the Hamiltonian has a finite range $\delta$, which means that there exists $\delta>0$ such that $[H_p,A]=0$ if $A$ is an observable localized at $q\in\Lambda$ and $|p-q|>\delta$. Therefore $[H_p,H_q]=0$ if  $|p-q|>2\delta$. We also assume that the operators $H_p$ are uniformly bounded: there exists $C>0$ such that $||H_p||<C$ for all $p\in\Lambda$.

Since we are interested in the thermodynamic limit, the subset $\Lambda$ is assumed infinite, but without accumulation points. Since we will be studying the net energy current through a section of a system, we assume that $\Lambda$ is compact in all but one direction with a coordinate $x\in\RR$. Then we may lump together all points at a particular $x\in\RR$ and regard our system as one-dimensional. From now on we focus on 1d lattice systems, with the lattice $\Lambda\subset\RR$ and a finite-range Hamiltonian 
$H=\sum_{x\in\Lambda} H_x$. 

We are interested in Gibbs states  at temperature $T=1/\beta$. We assume that the state is clustering, i.e. correlators of local operators 
$\langle A B\rangle-\langle A\rangle\langle B\rangle $
approach zero as $L_{AB}=\dist(\supp(A),\supp(B))\ra \infty.$ We also assume that the Kubo pairing 
\begin{equation}
\langle\langle A; B\rangle\rangle=\frac{1}{\beta}\int_0^\beta du\, \langle A(-iu)B\rangle-\langle A\rangle\langle B\rangle
\end{equation}
of any two local operators $A$ and $B$ 
decays at least as $L_{AB}^{-(1+\eps)}$ for some $\eps>0$. Here $A(t)=e^{iHt}A e^{-iHt}$. The Kubo pairing arises when studying the response to an infinitesimal perturbation $H\ra H+\lambda B$ \cite{Kubo}. Then the change in the equilibrium expectation value of $A$ is
\begin{equation}
\delta \langle A\rangle =\langle \delta A\rangle-\lambda\beta \langle\langle A; B\rangle\rangle .
\end{equation}
Here the first term accounts for the possible dependence of the observable $A$ on $\lambda$, while the second terms is due to the change of the equilibrium state.
Thus, up to a factor $\beta$, the Kubo pairing of local operators is the same as a generalized susceptibility for local perturbations.
The assumption that the Kubo pairing decays  faster than $1/L_{AB}$ ensures that perturbations of the form
$\sum_x B_x$
where $B_x$ is finite-range, and $||B_x||$ is uniformly bounded, lead to a well-defined change in the expectation values of all local observables. 

These decay assumptions are likely true for any positive temperature. Correlators of local observables decay exponentially away from phase transitions. One also expects the generalized susceptibilities for uniform perturbations to be finite away from phase transitions, although we are not aware of a proof. Since we are considering 1d systems, we do not expect any phase transitions at positive temperatures. Zero-temperature states can then be treated as $T\ra 0$ limits of Gibbs states.

\paragraph{Energy currents in lattice systems}\label{sec:latticecurrents}

By definition, the energy at a site $x\in\Lambda$ is $H_x$, and its time derivative is
\begin{equation}\label{timederivative}
\frac{dH_x}{dt}=i[H,H_x]=i\sum_{y\in\Lambda} [H_y,H_x].
\end{equation}
Following \cite{Kitaev} (see also  \cite{Mahan}), we define the energy current from site $y$ to site $x$ to be
\begin{equation}
J^E_{xy}=i[H_y,H_x].
\end{equation}
Then eq. (\ref{timederivative}) takes the form of a local conservation equation:
\begin{equation}
\frac{d H_x}{dt}=\sum_{y\in\Lambda} J^E_{xy}.
\end{equation}

Any $a\in\RR\backslash\Lambda$ divides $\Lambda$ into two parts: $\Lambda=\Lambda_+(a)\cup\Lambda_-(a)$, where $\Lambda_+(a)$ (resp. $\Lambda_-(a)$) is defined by the condition $x>a$ (resp. $x<a$). The net current from $\Lambda_-(a)$ to $\Lambda_+(a)$ is
\begin{equation}
J^{E}(a)=\sum_{x>a,y<a} J^{E}_{xy}. 
\end{equation}
For any $a,b\notin\Lambda$ and $b>a$ we have
\begin{equation}\label{conservationeq}
J^{E}(b)-J^{E}(a)=-\sum_{a<x<b} i[H,H_x].
\end{equation}
Since $\langle [H,A]\rangle=0$ for any local observable $A$, we get that $\langle J^E(a)\rangle$ is independent of $a$. 

Consider an infinitesimal variation of the Hamiltonian $\delta H$ such that $\delta H_x=0$ for sufficiently large positive $x$. Then
\begin{equation}\label{currentvariation}
\delta \langle J^E(a)\rangle=\langle \delta J^E(a)\rangle -\beta\langle\langle J^E(a);\delta H\rangle\rangle.
\end{equation}
Pick an $R>0$ such that $a+R\notin\Lambda$. Using the equation (\ref{conservationeq}) and the property of the Kubo pairing 
\begin{equation}
\langle\langle [H,A];B\rangle\rangle=\frac{1}{\beta}\langle[B,A]\rangle,
\end{equation}
the second term in (\ref{currentvariation}) can be written as
\begin{equation}
-\beta\langle\langle J^E(a);\delta H\rangle\rangle =-\beta\langle\langle J^E(a+R);\delta H\rangle\rangle-\sum_{a<x<a+R}\langle \,i [\delta H, H_x]\,\rangle .
\end{equation}
On the other hand, varying eq. (\ref{conservationeq}) we can re-write the first term in eq. (\ref{currentvariation}) as
\begin{equation}
\langle \delta J^E(a)\rangle=\langle \delta J^E(a+R)\rangle+\sum_{a<x<a+R}\langle \, i [\delta H,H_x]\,\rangle .
\end{equation}
Hence
\begin{equation}
\delta\langle J^E(a)\rangle=\langle \delta J^E(a+R)\rangle-\beta\langle\langle J^E(a+R);\delta H\rangle\rangle .
\end{equation}
Now let us take the limit $R\ra +\infty$. The first term is zero for sufficiently large $R$ since $\delta H_x=0$ for sufficiently large positive $x$, and both $H_x$ and $\delta H_x$ are assumed to have finite support, for all $x\in\Lambda$. Using the assumed decay of the Kubo pairing, the second term can be estimated to be no larger than $C/R^\eps$ for some $C>0$ and thus goes to zero for $R\ra +\infty$. Hence $\delta\langle J^E(a)\rangle=0$.

A similar argument shows that $\delta\langle J^E(a)\rangle=0$ for variations of $H$ which vanish for sufficiently large negative $x$. Now, any deformation of the Hamiltonian can be decomposed a sum of two deformations: one vanishing for $x\gg 0$ and one vanishing for $x\ll 0$. Linearity of response to infinitesimal deformation implies that variation of the current expectation value is the sum of variations corresponding to the two deformations. Since each of them vanishes, we conclude that $\delta\langle J^E(a)\rangle=0$ for arbitrary variations of $H$ within the allowed class.

Now we consider the temperature dependence of the net energy current. Re-scaling simultaneously the temperature $T\mapsto\lambda T$ and the Hamiltonian $H\mapsto \lambda H$ leaves the state unchanged, thus for any observable $A$ which does not depend explicitly on $T$ or $H$ we have
\begin{equation}
\left(T\frac{d}{dT}+\lambda\frac{d}{d\lambda}\right)\langle A\rangle_{\lambda}=0,
\end{equation}
where $\langle A\rangle_\lambda$ denotes the average over a Gibbs state with a Hamiltonian $\lambda H$ and temperature $T$. 
More generally, if $A$ is multiplied by $\lambda^p$ under $H\mapsto \lambda H$, then
\begin{equation}
\left(T\frac{d}{dT}+\lambda\frac{d}{d\lambda}\right)\left\langle \frac{A}{T^p}\right\rangle_\lambda=0.
\end{equation}
The energy current $J^E$  has  $p=2$. On the other hand, since re-scaling the Hamiltonian by a constant factor is an allowed deformation, we have
\begin{equation}
\frac{d}{d\lambda}\langle J^{E}(a)\rangle_\lambda=0.
\end{equation}
Therefore
\begin{equation}
\langle J^E(a)\rangle=CT^2,
\end{equation}
where $C$ is some constant which is unchanged under all  allowed variations of the Hamiltonian. 

Finally, let us assume that our state can be continuously connected to the maximally mixed state $T=\infty$. Then the above temperature dependence is incompatible with the fact that the operators $J^{E}(a)$ are bounded, unless $C=0$. Thus the net energy  current vanishes.

\paragraph{Energy currents in 
particle systems}\label{sec:particles}

There is a well-known difficulty with defining a local energy current in systems of particles with a potential interaction. It  is related to the non-locality of the potential interaction. One way of dealing with this difficulty involves a formal expansion of the potential $V({\bf x}-{\bf y})$ into an infinite sum of zero-range potentials (the Dirac delta-function $\delta({\bf x}-{\bf y})$ and its derivatives) \cite{Kirkwood}. If desired, one can smear the delta-function into a Gaussian, but the infinite sum remains \cite{Hardy}. Such an energy current is local only up to exponentially small ``tails.'' For 1d or quasi-1d systems with a finite-range potential there is an alternative approach which avoids both infinite series and ``tails'': one can define the energy density and the energy current which are local only in one dimension. This is sufficient for our purposes. To simplify notation, we will only discuss the strictly 1d case, but the modifications to the quasi-1d case are minor.

The second-quantized Hamiltonian has the form
\begin{equation}
H=\int dx\ k(x)+\int dx\ \rho(x) W(x)+\frac12\int dx dy\  \rho(x)\rho(y) V(x,y),
\end{equation}
where $k(x)$ is the usual kinetic energy density operator,
\begin{equation}
k(x)=\frac{1}{2m}\partial_x\psi^\dagger(x)\partial_x\psi(x),
\end{equation}
$\rho(x)=\psi^\dagger(x)\psi(x)$ is the particle density operator, $W(x)$ is the external potential, and $V(x,y)=V(y,x)$ is the pairwise interaction potential. We define the potential energy density as 
\begin{equation}
\pi(x)=W(x)\rho(x)+\frac12\rho(x)\int V(x,y)\rho(y) dy,
\end{equation}
and the total energy density as $h(x)=k(x)+\pi(x).$
To find the energy current $j^E(x)$, we need to solve the conservation equation
\begin{equation}
i[H,h(x)]=-\partial_x j^E(x).
\end{equation}
When computing the commutator on the left, the following identities are useful:
\begin{equation}
[\rho(x),\rho(y)]=0,\quad [\rho(x), j^Q(y)]=\frac{i}{m} \rho(y)\partial_y\delta(x-y), \quad i[k(x),\rho(y)]=j^Q(x)\,\partial_x \delta(x-y),
\end{equation}
where $j^Q=\frac{-i}{2m} (\psi^\dagger\partial_x \psi-(\partial_x\psi^\dagger)\psi)$ is the particle-number current. A solution has the form 
$j^E(x)=j_k^E(x)+j^E_\pi(x)$, where 
\begin{equation}
j^E_k(x)=\frac{-i}{4m^2} \left(\partial_x\psi^\dagger(x)\partial_x^2\psi(x)-\partial_x^2\psi^\dagger(x)\partial_x\psi(x)\right),
\end{equation}
and
\begin{multline}
j^E_\pi(x)=j^Q(x) W(x)+j^Q(x)\int V(x,y)\rho(y) dy+\frac{i}{2m} \rho(x) \left(\partial_x V(x,y)\right)\vert_{y=x}+\\
+\frac{1}{2}\int_{z<x<y} \left(\partial_y j^Q(y)\rho(z)-\partial_z j^Q(z)\rho(y)\right) V(y,z) dy dz.
\end{multline}
One can check that the energy current is Hermitian. Note that if the potential $V(x,y)$ has range $\delta$, i.e. vanishes whenever $|x-y|\geq\delta$, all terms in $j^E_\pi(x)$ are quasi-local: they commute with all local observables whose support is farther from $x$ than $\delta$. It is important for what follows that a quasi-local energy current can be constructed for an arbitrary symmetric finite-range potential $V(x,y)$. 

For any bounded function $\phi:\RR\ra \RR$ we can consider a modified potential $V_\phi(x,y)=\phi(x)\phi(y) V(x,y)$, which is also symmetric and finite-range. If $\phi(x)$ is small in some region of space, particle interactions are suppressed there. We claim that the energy current $\langle j^E(a)\rangle$ does not change as one varies $\phi$, provided the Kubo pairings of local operators decay at least as $1/L^{1+\eps}$. Indeed, consider an arbitrary infinitesimal variation of $\phi(x)$. It can be decomposed into a sum of two contributions: one vanishing for $x\ll 0$ and another one vanishing for $x\gg 0$. It is sufficient to show that the the energy current is unchanged under the two separately. Let us consider a variation of $\phi$ which vanishes for $x\gg 0$. As in the previous section, using the conservation equation and its variation we find:
\begin{equation}
\delta\langle j^E(a)\rangle=\langle\delta j^E(a+R)\rangle-\beta\langle\langle j^E(a+R);\delta H\rangle\rangle,
\end{equation}
where $R$ is arbitrary. Taking the limit $R\ra +\infty$, we conclude that $\langle j^E(a)\rangle $ is unchanged under arbitrary infinitesimal variations of $\phi$ which vanish for $x\gg 0$. An identical argument shows that $\langle j^E(a)\rangle $ is unchanged under arbitrary infinitesimal variations of $\phi$ which vanish for $x\ll 0$. Linearity of response then implies that $\langle j^E(a)\rangle$ is unchanged under arbitrary bounded variations of $\phi$.

Now let us take a constant $\phi=1$ and decrease it to $0$ (while keeping the temperature fixed). Unless one passes through a phase transition with divergent susceptibilities, $\langle j^E(a)\rangle $ is unchanged. Since it vanishes when $V(x,y)=0$, it must also be zero for the initial potential $V(x,y)$. It is widely believed that finite-temperature phase transitions cannot occur in systems of 1d particles with finite-range potential interactions. Assuming this, we proved that the equilibrium energy current vanishes for all $T>0$ and therefore also for $T=0$.

\paragraph{$U(1)$ currents in continuous systems}

In this section we discuss why Bloch's result does not apply to some continuous systems, like chiral 1+1d CFTs, but applies to others, like  non-relativistic particles. 

Consider a continuous system in $n$ spatial dimensions with a Hamiltonian
$H=\int h({\bf x}) d^n x$.
We assume time-translation symmetry but not necessarily spatial translation symmetry. The space is assumed to have the form $\RR\times W$, where $W$ is compact. The energy density $h({\bf x})$ is assumed to be quasi-local, in the sense that there exists a $\delta>0$ such that for any strictly local $A({\bf x})$ (i.e. a function of fields and their derivatives at a point ${\bf x}$) we have 
$
[h({\bf y}),A({\bf x})]=0
$
whenever $|{\bf x}-{\bf y}|>\delta .$ Both local field theories (whether Lorentz-invariant or not) and non-relativistic particles with finite-range interactions obey this.

We assume that the $U(1)$ generator $Q$ is $Q=\int \rho({\bf x}) d^n x$ where $\rho$ is a local operator, and that there exists a quasi-local $U(1)$ current ${\bf j}^Q({\bf x})$ satisfying 
\begin{equation}\label{ucurrent}
i[H,\rho({\bf x})]=-\nabla\cdot {\bf j}^Q({\bf x}). 
\end{equation}
This condition is satisfied for local field theories as well as for systems of non-relativistic particles.

Suppose we can promote $U(1)$ symmetry to a local $U(1)$ symmetry with generators
\begin{equation}
Q_f=\int \rho({\bf x}) f({\bf x}) d^n x,
\end{equation}
where $f({\bf x})$ is an arbitrary function. Requiring $[Q_f,Q_g]=0$ for all $f,g$ we get 
\begin{equation}\label{chargedensitycomm}
[\rho({\bf x}),\rho({\bf y})]=0.
\end{equation}
Using (\ref{chargedensitycomm}) we can deduce that the net $U(1)$ current, if present, cannot depend on the chemical potential $\mu$. 
Indeed, consider an infinitesimal deformation of the Hamiltonian of the form
\begin{equation}\label{defH}
\delta H=\int f({\bf x}) \rho({\bf x}) d^n x.
\end{equation}
The condition (\ref{chargedensitycomm}) implies that the current is undeformed, $\delta {\bf j}^Q=0$, regardless of $f({\bf x})$. Following the same procedure as above, we find the change in the expectation value of $J^Q(a)=\int_W j_1^Q(a,w) d^{n-1} w$:
\begin{equation}
\delta \langle  J^Q(a) \rangle=-\beta \langle\langle J^Q(a+R); \int f({\bf y}) \rho({\bf y}) d^n y\rangle\rangle ,
\end{equation}
where $R$ is arbitrary. Writing a general bounded $f({\bf x})$ as a sum of two functions vanishing for $x\gg 0$ and $x\ll 0$ we argue as before that $\langle  J^Q(a) \rangle=0$. Then, taking $f$ to be constant, we deduce that $\langle J^Q(a)\rangle$ is independent of the chemical potential, provided we stay away from phase transitions. For non-relativistic particles, we can deform $\mu$ to $-\infty$ and get Bloch's result.

In 1+1d CFTs with a non-zero $k_R-k_L$ instead of  (\ref{chargedensitycomm}) one has $[\rho(x),\rho(y)]=-i c \delta'(x-y)$ where $c=(k_R-k_L)/2\pi$. Such $c$-number terms in the commutators are known as  Schwinger terms. Since the deformation (\ref{defH}) no longer commutes with $\rho(x)$, the current now depends on $f$. One finds $j^Q(x,f)=j^Q(x)+c f(x)\cdot 1$. Going through the same argument as above, one finds that $\langle j^Q(x,f)\rangle$ is independent of $f$. Setting $f(x)=-\mu$, we seem to find that the equilibrium current vanishes at arbitrary $\mu$, in contradiction with (\ref{chiralk}). However, the current which appears in (\ref{chiralk}) is not $j^Q(x,f)$ for $f(x)=-\mu$, but the undeformed current $j^Q(x)$. For a constant $f(x)$, the conservation equation can be satisfied without changing the current thanks to $[Q,\rho(x)]=0,$ and this is the standard choice in CFT. Taking into account the relation between $j^Q(x,f)$ and $j^Q(x),$ we indeed find (\ref{chiralk}). 

The discrepancy between $j^Q(x,f)$ and $j^Q(x)$ is a manifestation of $U(1)$ anomaly. One can think of the deformation (\ref{defH}) as coupling the theory to an external electric potential $\varphi=f(x)$. The current $j^Q(x,f)$ depends on $f(x)$ locally, but it is not invariant under the transformation $f(x)\mapsto f(x)+f_0$, which is a particular gauge transformation. To patch this up one can define a current $\tilde j^Q(x,f)=j^Q(x,f)-c f(a)\cdot 1,$ where $a$ is an arbitrary point. This current is gauge invariant and reduces to $j^Q(x)$ when $f(x)=-\mu$. However, $\tilde j^Q(x,f)$ is non-local. The conflict between gauge-invariance and locality is a manifestation of $U(1)$ anomaly.

\paragraph{Applications}

We showed that the equilibrium energy current vanishes both for infinitely-extended 1d lattice systems with finite-range interactions and quasi-1d systems of non-relativistic particles with finite-range potential interactions. The only assumption was the absence of phase transitions at positive temperatures, which is expected to hold for such systems. In view of eq.~(\ref{chiralc}), our result implies that such systems cannot flow to a 1+1d CFT with a nonzero $c_R-c_L$.

One subtlety in this argument is that the CFT energy current $T^{10}$ might not be the same as the infrared limit $j^E$ of the microscopic energy current. As any locally conserved quantity, stress-energy tensor $T^{\mu\nu}$ is not completely unique, and the freedom to redefine it might be important in order to ensure that it is symmetric and traceless. However, since both $T^{00}$ and the microscopic energy density $h(x)$ must integrate to the same low-energy Hamiltonian, they can differ at most by a total derivative: $T^{00}=h+\partial_x O,$ where $O$ is a local operator. Then the CFT energy current $T^{10}$ is related to $j^E$ by $T^{10}=j^E-\partial_t O$. Hence the equilibrium expectation values of $T^{10}$ and $j^E$ are the same, and the vanishing of $\langle j^E\rangle$ implies $\langle T^{10}\rangle =0$ and $c_R-c_L=0.$

It is well-known that a nonzero $c_R-c_L$ may appear in 1+1d CFTs describing the gapless edge of a gapped 2d system. The above result shows that $c_R-c_L$ is determined by the bulk properties of the 2d material and does  not depend on the edge. Indeed, we may consider a strip of the 2d phase bounded by two different edges (with opposite orientations) as a 1d material, and then the above result shows that $c_R-c_L$ must be equal for the two edges. This is not surprising since $c_R-c_L$ is related to the bulk thermal Hall conductance. The same comments apply, {\it mutatis mutandis}, to $k_R-k_L$ and the Hall conductance. 

The vanishing of the net $U(1)$ current is implicitly assumed in the definition of  magnetization. Usually, one says that since $\nabla \cdot \langle {\bf j}^Q({\bf x})\rangle=0$ in an equilibrium state, one can define the magnetization by the equation
$
\nabla\times {\bf M^Q}({\bf x})=\langle {\bf j}^Q({\bf x})\rangle.
$
If the net current in some direction were nonzero, the magnetization so defined would be either multi-valued (if the direction is periodically identified) or would grow linearly with distance. 
In either case, it could not be regarded as a local property of the material. Bloch's theorem shows that the magnetization is well-defined. An analogous quantity for energy currents (``energy magnetization'') is of importance in the theory of the thermal Hall effect \cite{Halperinetal,Niuetal}. Our results on the vanishing of the net energy current justify the existence of energy magnetization in a wide variety of situations.

One final remark is that the vanishing of   $U(1)$ and energy currents strictly applies to infinite systems in equilibrium. In a large but finite quasi-1d system, like a macroscopic ring, there can be a non-vanishing $U(1)$ or energy current in equilibrium. However, it must go to zero when the size of the system goes to infinity.

\end{document}